\documentclass[12pt, a4paper]{article}

\begin{document}

\title{Quantum comparison machines \\
with one-sided error}
\author{Chiu Fan Lee\thanks{c.lee1@physics.ox.ac.uk}
\ and \ Neil F. Johnson\thanks{n.johnson@physics.ox.ac.uk}
\\
\\ Centre for Quantum Computation and Physics Department \\ Clarendon
Laboratory, Oxford University \\ Parks Road, Oxford OX1 3PU, U.K.}

\maketitle

\abstract{It is always possible to decide, with one-sided error,
whether two quantum states are the same under a specific unitary
transformation. However we
show here that it is {\em impossible} to do so if the transformation is
anti-linear
and non-singular. This result implies that unitary and anti-unitary
operations exist on an
unequal footing in quantum information theory. }

\newpage

The formalism of quantum mechanics has given rise to many theorems
concerning
`impossible machines'. For example, many quantum machines such
as quantum copiers \cite{WZ82} and quantum erasers \cite{PB00} are
impossible. However these operations can be
performed approximately
\cite{clone, We98}. Furthermore,
Galv\~{a}o and Hardy have shown that approximate cloning can actually enhance
the performance of some computations \cite{GH00}.
It is therefore very important to
understand the limitations of operations which are `approximately
impossible'.

Here we consider quantum decision machines with a one-sided error
probability. These
machines output the answer {\tt YES} or {\tt NO}, 
however the probability of error is
non-zero for
one of the two possible outputs. Although this concept  sounds quite
theoretical, practical
examples of such one-sided machines are not unfamiliar to physicists. In
particular the interaction-free measurement scheme
introduced by Elitzur
and Vaidman in Ref. \cite{EV93}
is inherently a decision machine with one-sided error.
In this  article, we study the fundamental
limit of a subclass
of quantum decision machines, which we call comparison machines. We show
that although it is
always possible to decide---with one-sided error---whether two states are
the same under a
specific unitary transformation, it is impossible to do so if the
transformation is anti-linear
and non-singular. The present work therefore demonstrates the unparallel
roles of unitary and
anti-unitary transformations in quantum information theory.

Without loss of generality, we
assume that the answer {\tt YES} or {\tt NO} 
is given by a measurement on the first
qubit of the
output state, and that
$|0\rangle$  corresponds to {\tt YES} while $|1\rangle$ corresponds to {\tt NO}. 
An immediate
example of a decision machine is the SWAP-test discussed in Ref.~\cite{BCW01}. Given
two physical systems in unknown states $|\phi\rangle $ and
$|\psi\rangle$, it is impossible to tell whether they are identical or 
not---however
Buhrman {\it et al} have shown in Ref. \cite{BCW01} that
it is indeed possible to do so in the presence of some one-sided
error. Explicitly, if $|\phi\rangle =|\psi\rangle$, then
the machine will always say {\tt YES}. However if $|\langle \phi|\psi\rangle
|=\delta<1$,  then the machine says {\tt YES} with probability $(1+\delta^2)/2$.
This SWAP-test was also utilized by Gottesman and
Chuang to construct their digital signature scheme \cite{GC01}.

A further motivation for studying such a comparison machine, is that
the output is a physical observable. For instance, we may use the
SWAP-test to determine the payoff to Clare (the cloner) in Werner's cloning
game \cite{We98} instead of the fidelity.
For example, we may
assign a payoff of 1 to Clare if her clones pass the test, and $-1$
otherwise. As the SWAP-test depends only on the fidelity of the two states,
in terms of expected payoff, this game and Werner's original one
are equivalent.
However the game is now physically implementable,
whereas Werner's orignal game was not because there is no physical operation
to compute the fidelity of two states. Furthermore, with the SWAP-test
adopted as the gauge for payoff, the game formalism and
the Quantum Minimax Theorem developed in Ref. \cite{LJ02t} can be applied to
show that 
Werner's cloning operation is indeed
the strategy at the Nash equilibrium \cite{LJ02n}.

Given two states $|\phi\rangle$ and $|\psi\rangle$,
a more general version of the SWAP-test is to test, with
some one-sided error probability, whether $|\psi\rangle
= K|\phi\rangle$ for some arbitrary operation $K$. If $K$ is
unitary, this is made trivially
possible using the SWAP-test on $K|\phi\rangle$ and $|\psi\rangle$.
On the other hand, we now show that such a device is not possible
if $K$ is anti-linear and non-singular. 
Given an arbitrary two-level state: $a|0\rangle + b|1\rangle$, the state 
orthogonal to it is $\bar{a}|1\rangle - \bar{b}|0\rangle$, so an
anti-linear map is needed to transform an arbitrary two-level
state to its orthogonal state. Therefore, our result
immediately rules out the
possibility of deciding whether two arbitrary states are orthogonal with
one-sided
error probability. Our result suggests that it is much more stringent to
test the
orthogonality condition than to test the identity condition. A complementary
idea was
discussed in Ref. \cite{GP99} where
Gisin and Popescu showed that
antiparallel pairs of states encode more
information than parallel pairs of states. Our result also contrasts
with the many proofs of impossibility covering both unitary and anti-unitary
maps \cite{Pa01}, and hence
demonstrates concretely the
unparallel roles of unitary and anti-unitary transformations in quantum
information theory.

We now give the proof of the statement, noting that the proof
only relies on the linearity assumption of quantum mechanics.
Given an arbitrary state $|\phi\rangle$ of particle 1 and 
another arbitrary state $|\psi\rangle$ of particle 2,
let $M$ be a machine capable of deciding whether $|\psi\rangle =
K|\phi\rangle$ with some one-sided error probability. We now denote
$K|0\rangle$ by $ |\phi_0 \rangle$ and
$K|1\rangle$ by $ |\phi_1 \rangle$. We first suppose that
the machine says {\tt NO} ($
|1\rangle$) with
certainty, but it errs when the answer is {\tt YES} ($|0\rangle$):
\begin{eqnarray*}
M: |0\rangle_1 |\phi_0\rangle_2 |ancillar\rangle &\mapsto &
a_{00}|0\rangle |A_{00}\rangle + b_{00}|1\rangle |B_{00}\rangle \\
|1\rangle_1 |\phi_1\rangle_2 |ancillar\rangle &\mapsto &
a_{11}|0\rangle |A_{11}\rangle + b_{11}|1\rangle |B_{11}\rangle \\
|0\rangle_1 |\phi_1\rangle_2
|ancillar\rangle &\mapsto & |1\rangle |B_{01}\rangle \\
|1\rangle_1 |\phi_0\rangle_2
|ancillar\rangle &\mapsto & |1\rangle |B_{10}\rangle.
\end{eqnarray*}
However, since $K$ is non-singular,
$|\phi_1\rangle \neq K(\frac{1}{\sqrt{2}}(|0\rangle +|1\rangle))$, hence $
M(\frac{1}{\sqrt{2}}(|0\rangle_1 +|1\rangle_1)|\phi_1\rangle_2)=|1\rangle |{\cal
B}\rangle$
for some state $|{\cal B}\rangle$.
This implies that $a_{11}=0$. Similarly one can show that $a_{00}=0$,
hence the machine is trivial.

We next suppose that the machine says {\tt YES} with
certainty, but errs when it says {\tt NO}. In this case
\begin{eqnarray*}
M:|0\rangle_1 |\phi_0\rangle_2
|ancillar\rangle &\mapsto & |0\rangle |A_{00}\rangle \\
|1\rangle_1 |\phi_1\rangle_2
|ancillar\rangle &\mapsto & |0\rangle |A_{11}\rangle \\
|0\rangle_1 |\phi_1\rangle_2 |ancillar\rangle &\mapsto &
a_{01}|0\rangle |A_{01}\rangle + b_{01}|1\rangle |B_{01}\rangle \\
|1\rangle_1 |\phi_0\rangle_2 |ancillar\rangle &\mapsto &
a_{10}|0\rangle |A_{10}\rangle + b_{10}|1\rangle |B_{10}\rangle
\end{eqnarray*}
Using linearity, $M(\frac{1}{2}(|0\rangle_1 +|1\rangle_1)
(|\phi_0\rangle_2 +|\phi_1\rangle_2))=\frac{1}{2}|0\rangle(|A_{00}\rangle+
a_{01}|A_{01}\rangle +
a_{10}|A_{10}\rangle+|A_{11}\rangle )+\frac{1}{2}|1\rangle(
b_{01}|B_{01}\rangle+b_{10}|B_{10}\rangle)$. However since
$K(|0\rangle +|1\rangle)=(|\phi_0\rangle +|\phi_1\rangle)$,
this implies $b_{01}|B_{01}\rangle +b_{10}|B_{10}\rangle=0$.
On the other hand, 
$M(\frac{1}{2}(|0\rangle_1 +i|1\rangle_1)
(|\phi_0\rangle_2 -i|\phi_1\rangle_2))=\frac{1}{2}|0\rangle(|A_{00}\rangle+
ia_{10}|A_{10}\rangle -i
a_{01}|A_{01}\rangle+|A_{11}\rangle)+\frac{i}{2}|1\rangle(
b_{10}|B_{10}\rangle-b_{01}
|B_{01}\rangle)$. Again since
$K(|0\rangle +i|1\rangle)=(|\phi_0\rangle -i|\phi_1\rangle)$
by anti-linearity,
we have $b_{01}|B_{01}\rangle -b_{10}|B_{10}\rangle=0$. Therefore,
the machine always outputs {\tt YES}.
This completes the proof of impossibility.

In conclusion, we have ruled out any physical devices that would decide
whether two states are the same under a non-singular anti-linear
transformation.
As a result, unitary and anti-unitary operations are placed on an {\em
unequal} 
footing.  This result suggests that further investigation is needed to
clarify the boundary of feasible operations in quantum information
theory.

\vskip0.5in CFL thanks NSERC (Canada), ORS (U.K.) and
the Clarendon Fund (Oxford) for financial support.

\end{document}